\documentclass{article}
\usepackage{ace,amsmath,graphicx,url,times,subfigure}
\usepackage{cmdtrack}

\pdfoutput = 1
\title{PSD estimation in Beamspace for Estimating Direct-to-Reverberant Ratio from A Reverberant Speech Signal}
\twoauthors
  {Yusuke Hioka}
    {University of Auckland\\
Department of Mechanical Engineering\\
Private Bag 92019, Auckland 1142, New Zealand \\
     yusuke.hioka@ieee.org}
  {Kenta Niwa}
    {NTT Corporation \\
     NTT Media Intelligence Laboratories\\
      3-9-11 Midori-cho, Musashino,\\ Tokyo 180-8585, Japan \\
     niwa.kenta@lab.ntt.co.jp}

\begin{document}

\ninept
\maketitle

\begin{sloppy}

\begin{abstract}
A method for estimation of direct-to-reverberant ratio (DRR) using a microphone array is proposed. The proposed method estimates the power spectral density (PSD) of the direct sound and the reverberation using the algorithm \textit{PSD estimation in beamspace} with a microphone array and calculates the DRR of the observed signal. The speech corpus of the ACE (Acoustic Characterisation of Environments) Challenge was utilised for evaluating the practical feasibility of the proposed method. The experimental results revealed that the proposed method was able to effectively estimate the DRR from a recording of a reverberant speech signal which included various environmental noise.

\end{abstract}


\section{Introduction}
\label{sec:intro}
Estimation of direct-to-reverberant ratio (DRR) has been attracting interests of researchers in audio and acoustic signal processing since the DRR is one of the important parameters in room acoustics which specifies the acoustic characteristics of a reverberant enclosure \cite{absorbtionrate}. The estimated DRR can also be utilised in various audio signal processing applications such as dereverberation \cite{dereverberation_book}, source distance estimation \cite{Lu2010,Vesa2007,Vesa2009,HIOKA2011IEEE}, emphasising speech located at a particular distance \cite{HIOKA2010ICASSP}, and spatial audio coding \cite{laitinen2012AES}. 

The room impulse response (RIR) used to be measured to calculate the DRR of a reverberant enclosure. However the measurement of the RIR was actually a hinderance for application users since special equipment and software are needed, which motivated researchers to estimate the DRR in \textit{blind} manner, i.e. without RIR measurement. Since the propagation properties of the direct sound and the reverberation are quite different, several recent studies utilised more than one microphones (i.e. a microphone array) for the blind DRR estimation problem \cite{Larsen2003}. Some recent studies achieved blind DRR estimation by introducing a model for the coherence of direct sound and reverberation between two microphones \cite{Vesa2007,Vesa2009,Jeub2011EUSIPCO,Theirgart2012ICASSP} since the magnitude-squared coherence (MSC) is a quantity related to the DRR \cite{Vesa2007,Vesa2009}. The authors also proposed a method based on a similar strategy but using more than two microphones \cite{HIOKA2011IEEE}. The method introduced a model for the spatial correlation matrix of a microphone array signal recorded in a reverberant environment which contains the coherence information between each pair of microphones. The method estimated the power spectral density (PSD) \cite{PSD} of the direct sound and the reverberation simultaneously using the introduced model. 

Besides the introduction of the model another novel aspect in this method was in the calculation of the DRR from the estimated PSD of the direct sound and the reverberation. The authors later proposed another DRR estimation method which estimates the PSD using two different beamformers whose beampatterns have an identical shape \cite{hioka_2012IWAENC}. It should be noted that these previous methods are \textit{semi-blind} methods because they assume the direction of the sound source is known \textit{a priori}. Another similar approach proposed by Thiergart \textit{et al.} also estimates the PSD for deriving the DRR \cite{Thiergart2014}, which uses multiple directional microphones rather than a microphone array. 

In this study another PSD based blind DRR estimation method is proposed. The proposed method utilises an algorithm: \textit{PSD estimation in beamspace} \cite{Hioka_ASLP2013,Hioka_IWAENC2014}, which was originally invented for sound source separation problems by the authors, to estimate the PSD of the direct sound and the reverberation. The method was evaluated on the speech corpus of the ACE (Acoustic Characterisation of Environments) Challenge \cite{Eaton2015a} and its performance was compared to that of the authors' previous studies \cite{HIOKA2011IEEE,hioka_2012IWAENC}. Since the problem specified by the ACE Challenge was a \textit{fully} blind problem (i.e. no prior information about the source direction is provided), the methods were tested after being combined with conventional direction-of-arrival (DOA) estimation and voice activity detection (VAD) methods. Presenting the performance of the authors' methods under a fully blind problem is another contribution of this paper.

\section{Problem setup}\label{sec:problem}
\subsection{Sound propagation in reverberant environment}

An impulse response from a sound source to a microphone (i.e. the sound propagation path) in a reverberant environment is generally described by a model which consists of the direct sound, early reflections, and late reverberation \cite{absorbtionrate}. However as shown in Fig. \ref{fig:impulse_response.pdf} in this study the impulse response is simply modelled by the two components, the direct sound and the reverberation, where the latter includes both the early reflections and the late reverberation. Given the impulse response in the frequency domain (i.e. the transfer function) is denoted by $H(\omega)$ where $\omega$ is the frequency, the model is represented by
\begin{align}
H(\omega)=H_{\mathrm{D}}(\omega)+H_{\mathrm{R}}(\omega),\label{eq:transfer_function}
\end{align}
where $H_{\mathrm{D}}(\omega)$ and $H_{\mathrm{R}}(\omega)$ are the transfer functions of the direct sound and the reverberation, respectively. 
\begin{figure}[tb]
  \begin{center}
    \includegraphics[keepaspectratio=true,width=0.40\textwidth]{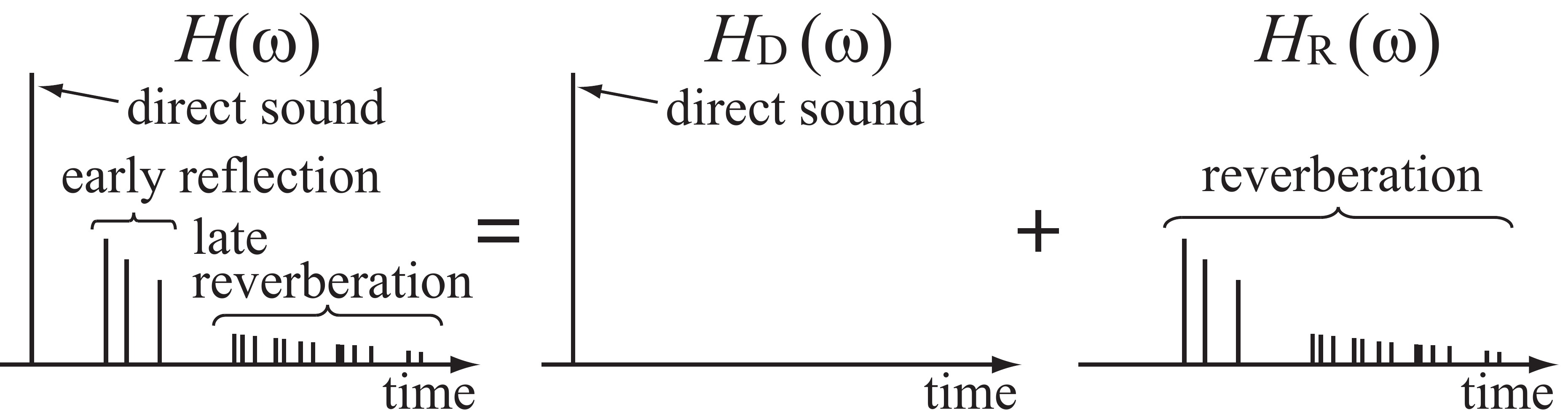}
  \end{center}
  \caption{Decomposition of impulse response between sound source and microphone.}%
\label{fig:impulse_response.pdf}
\end{figure}
The sound propagation paths of these two components are quite different. In general the direct sound is assumed to be spatially coherent, whereas the reverberation is assumed to be diffuse \cite{absorbtionrate}. In other words, as described in Fig. \ref{fig:isotropic_propagation.pdf}, the direct sound arrives into a microphone directly only from the source direction (directional). In contrast, the reverberation arrives from every direction with uniform energy distribution (isotropic). The proposed method focuses on the difference in these spatial properties of the sound propagation paths and segregate the direct sound and the reverberation using a microphone array. 
The rest of this section discusses the signal modelling for the microphone array observation and the output of the beamforming \cite{Johnson_arraybook}, which is used to estimate the PSD of the direct sound and the reverberation for calculating the DRR. 

In the following discussion a few assumptions are made: the source direction from the microphone array is known \textit{a priori} using any conventional direction of arrival (DOA) estimation techniques \cite{brandstein2001microphone}; the direct sound and the reverberation are mutually uncorrelated; and both direct sound and the reverberation are modelled by plane waves \cite{Johnson_arraybook}.
\begin{figure}[tb]
  \begin{center}
    \includegraphics[keepaspectratio=true,width=0.35\textwidth]{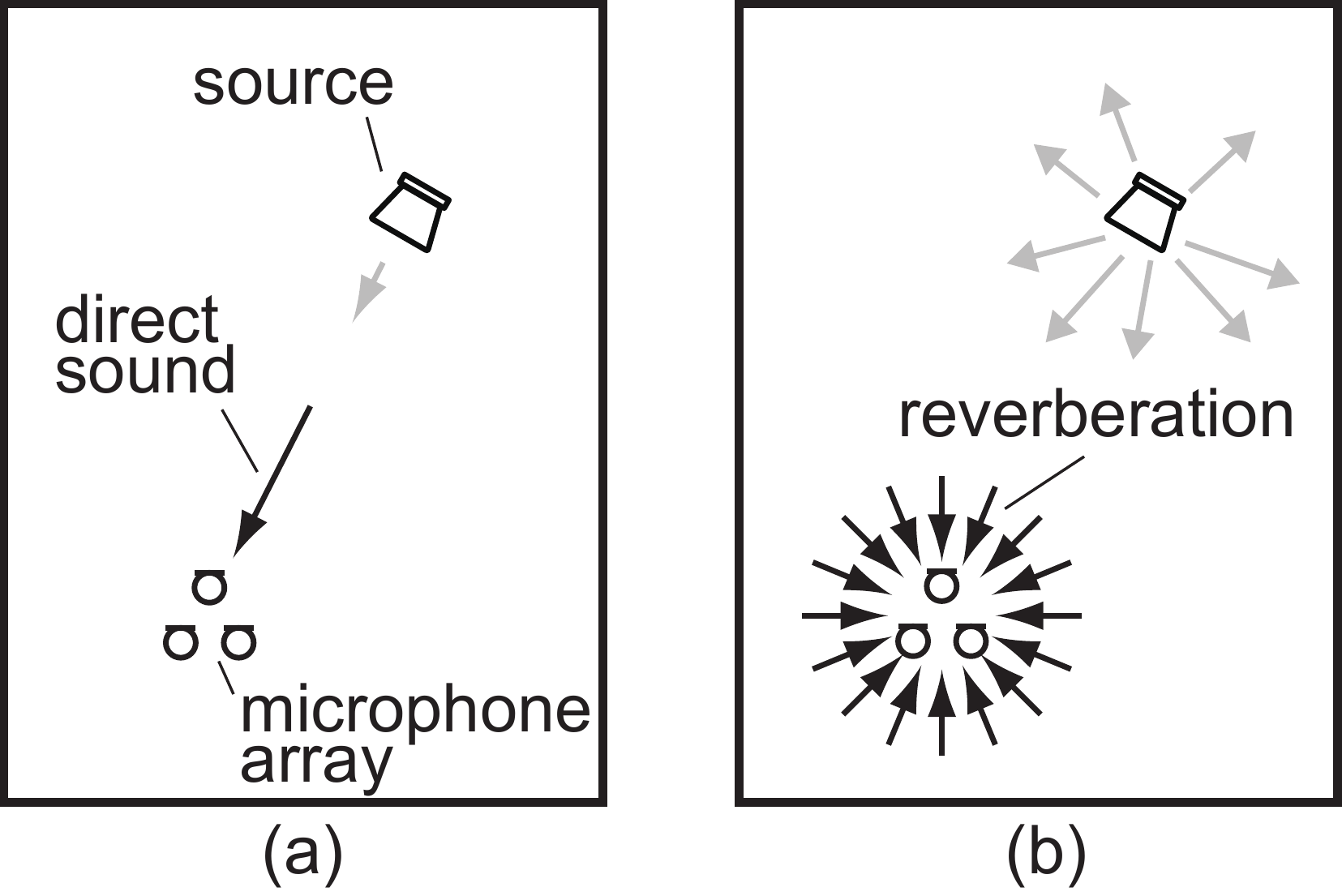}
  \end{center}
\vspace*{-1.0\baselineskip}
  \caption{Propagation path from sound source to microphone array in reverberant room: (a) direct sound, (b) reverberation.}
  \label{fig:isotropic_propagation.pdf}
\end{figure}
\subsection{Microphone array observation}\label{subsec:mic_observ}
Let $X^{(m)}(\omega,t)$ be the observed signal of the $m$-th microphone of an $M$-sensors microphone array in the time-frequency domain where $t$ is a frame index. Using the transfer function of the sound propagation path introduced in (\ref{eq:transfer_function}), $X^{(m)}(\omega,t)$ can be modelled by
\begin{align}
X^{(m)}(\omega,t)&:=\left(H_{\mathrm D}^{(m)}(\omega)+H_{\mathrm R}^{(m)}(\omega)\right)S(\omega,t),\label{eq:array_input}
\end{align}
where $S(\omega,t)$ is the spectrum of a sound source. 

The transfer functions in (\ref{eq:array_input}) are further decomposed into two components: the transfer function from the sound source to a reference point located close to the microphone array (e.g. the centre of the microphone array), and that from the reference point to each microphone. Since the signals are assumed to be plane waves, the latter transfer function can be approximated by the phase shift caused by the propagation delay. Namely $H_{\mathrm D}^{(m)}(\omega)$ and $H_{\mathrm R}^{(m)}(\omega)$ are expressed by
\begin{align}
H_{\mathrm D}^{(m)}(\omega)&=H_{\mathrm{Dref}}(\omega)e^{-j\omega\tau^{(m)}_{\Omega_{\mathrm{D}}}},\label{eq:TFunc_direct}\\
H_{\mathrm R}^{(m)}(\omega)&=\int_{\Omega}H_{\mathrm{Rref},\Omega}(\omega)e^{-j\omega\tau^{(m)}_{\Omega}}d\Omega,\label{eq:TFunc_reverb}
\end{align}
where $H_{\mathrm{Dref}}(\omega)$ and $H_{\mathrm{Rref},\Omega}(\omega)$ are the transfer functions from the sound source to the reference point with regard to the direct sound and the reverberation, respectively. $\tau_{\Omega}^{(m)}$ is the propagation delay from the reference point to the microphone $m$ when the sound wave is arriving from a solid angle $\Omega=\{\theta,\phi\}$, where $\theta$ and $\phi$ are the azimuth and the zenith angles, respectively ($\theta\in[0,2\pi)$, $\phi\in[0,\pi]$). Note that $\phi=\frac{\pi}{2}$ is equal to the planer in parallel to the plane of the microphone array, and $\Omega_{\mathrm{D}}$ denotes the source direction.

The observation vector of the microphone array is defined by 
\begin{align}
&\hspace{-2em}\mathbf{x}(\omega,t)=[X^{(1)}(\omega,t),\cdots,X^{(M)}(\omega,t)]^{T}\nonumber\\
&=\mathbf{a}_{\Omega_{\mathrm{D}}}(\omega)S_{\mathrm{D}}(\omega,t)+\int_{\Omega}\mathbf{a}_{\Omega}(\omega)S_{\mathrm{R},\Omega}(\omega,t)d\Omega,\label{eq:array_input_vector}
\end{align}
where $\mathbf{a}_{\Omega}(\omega)=[e^{-j\omega\tau^{(1)}_{\Omega}}, \cdots, e^{-j\omega\tau^{(M)}_{\Omega}}]^{T}$ is the steering vector \cite{Johnson_arraybook} for the angle $\Omega$, and 
\begin{align}
&S_{\mathrm{D}}(\omega,t)=H_{\mathrm{Dref}}(\omega)S(\omega,t),\\
&S_{\mathrm{R},\Omega}(\omega,t)=H_{\mathrm{Rref},\Omega}(\omega)S(\omega,t),
\end{align}
are the direct sound and the reverberation arriving from the angle $\Omega$ observed at the reference point, respectively. $T$ denotes the transpose of a vector or a matrix.
\subsection{Beamforming output}
Assume more than one arbitrary but different beamformers are applied to the microphone array observation $\mathbf{x}(\omega,t)$, then the output signal of the beamformer $l$ is represented by
 \begin{align}
Y_{\mathrm{BF},l}(\omega)&=\mathbf{w}_{l}^{H}(\omega)\mathbf{x}(\omega,t)\label{eq:BF_out_define}\\
&=\mathbf{w}_{l}^{H}(\omega)\mathbf{a}_{\Omega_{\mathrm{D}}}S_{\mathrm{D}}(\omega,t)\nonumber\\
&\hspace*{1.5em}+\mathbf{w}_{l}^{H}(\omega)\int_{\Omega}\mathbf{a}_{\Omega}S_{\mathrm{R},\Omega}(\omega,t)d\Omega,\label{eq:BF_out_true}
\end{align}  
where $\mathbf{w}_{l}(\omega)$ is the weight vector of the beamformer $l$ defined by
\begin{align}
\mathbf{w}_{l}(\omega)=[W_{l}^{(1)}(\omega),\cdots,W_{l}^{(M)}(\omega)]^{T}.
\end{align}

The PSD of the beamformer's output is then described by the summation of the PSD of the direct sound and the reverberation multiplied by the gain of the beamformer:
\begin{align}
\hspace*{-1.5em}P_{\mathrm{BF},l}(\omega)&=E[|Y_{\mathrm{BF},l}(\omega)|^2]_{t}\\
\hspace*{-1.5em}&\approx\mathbf{w}_{l}^{H}(\omega)\mathbf{a}_{\Omega_{\mathrm{D}}}(\omega)E[|S_{\mathrm{D}}(\omega,t)|^{2}]_{t}\mathbf{a}_{\Omega_{\mathrm{D}}}^{H}(\omega)\mathbf{w}_{l}(\omega)\nonumber\\
&\hspace{1.5em}+\mathbf{w}_{l}^{H}(\omega)\left\{\int_{\Omega}\mathbf{a}_{\Omega}(\omega)E[|S_{\mathrm{R},\Omega}(\omega,t)|^{2}]_{t}\right.\nonumber\\
&\hspace{10em}\left.\cdot~ \mathbf{a}_{\Omega}^{H}(\omega)d\Omega\right\}\mathbf{w}_{l}(\omega)\label{eq:BF_out_PSD_true}\\
\hspace*{-1.5em}&=G_{l,\Omega_{\mathrm{D}}}(\omega)P_{\mathrm{D}}(\omega)+\int_{\Omega}G_{l,\Omega}(\omega)P_{\mathrm{R},\Omega}(\omega)d\Omega,\label{eq:BF_out_PSD}
\end{align}
where $P_{\mathrm{D}}(\omega)$ and $P_{\mathrm{R},{\Omega}}(\omega)$ are the PSD of the direct sound and the reverberation observed at the reference point, respectively:
\begin{align}
P_{\mathrm{D}}(\omega)&=E[|S_{\mathrm{D}}(\omega,t)|^{2}]_{t},\\
P_{\mathrm{R},{\Omega}}(\omega)&=E[|S_{\mathrm{R},{\Omega}}(\omega,t)|^{2}]_{t}.
\end{align}
$E[\cdot]_{t}$ is the expectation over frames that can be approximated by the average of several frames, and $G_{l,\Omega}(\omega)$ is the gain of the beamformer for angle $\Omega$ defined by
\begin{align}
G_{l,\Omega}(\omega)=|\mathbf{w}_{l}^{H}(\omega)\mathbf{a}_{\Omega}(\omega)|^{2}.
\end{align}
In the derivation of (\ref{eq:BF_out_PSD}), the uncorrelatedness assumption for the relation between the direct sound and the reverberation, i.e. $E[S^{*}_{\mathrm{D}}(\omega,t)S_{\mathrm{R},\Omega}(\omega,t)]_{t}=0$, is utilised.

Since the isotropic propagation has been assumed for the reverberation, the PSD of the reverberation can be replaced by a constant value that holds for all $\Omega$, i.e.
\begin{align}
P_{\mathrm{R},\Omega}(\omega)=\overline{P}_{\mathrm{R}}(\omega)=\mathrm{const.}~~~~~~\forall \Omega.\label{eq:PSD_rev_constant}
\end{align}
Thus, the PSD of the beamformer output in (\ref{eq:BF_out_PSD}) becomes
\begin{align}
P_{\mathrm{BF},l}(\omega)=G_{l,\Omega_{\mathrm{D}}}(\omega)P_{\mathrm{D}}(\omega)+\overline{P}_{\mathrm{R}}(\omega)\int_{\Omega}G_{l,\Omega}(\omega)d\Omega .\label{eq:BF_out_PSD_isotropic}
\end{align}
\subsection{Calculating DRR from estimated PSD}
The DRR is commonly defined by the energy ratio of the impulse response for the direct sound to that for the reverberation \cite{Larsen2003}. According to Parseval's theorem and with some approximations, the DRR can also be calculated from the PSD of the direct sound and the reverberation observed at the reference point of the microphone array
\begin{align}
\Gamma[\mathrm{dB}]&:=10\log_{10}\left(\frac{\sum_{\omega}|H_{\mathrm{D}}(\omega)|^{2}}{\sum_{\omega}|H_{\mathrm{R},\Omega}(\omega)|^{2}}\right)\\
&\approx10\log_{10}\left(\frac{\sum_{\omega}P_{\mathrm{D}}(\omega)}{\sum_{\omega}\int_{\Omega}P_{\mathrm{R},\Omega}(\omega) d\Omega}\right)\\
&=10\log_{10}\left(\frac{\sum_{\omega}P_{\mathrm{D}}(\omega)}{4\pi\sum_{\omega}\overline{P}_{\mathrm{R}}(\omega)}\right)\label{eq:DRR}.
\end{align}
In the following section an algorithm to estimate the PSD, i.e. $P_{\mathrm{D}}(\omega)$ and $\overline{P}_{\mathrm{R}}(\omega)$, is introduced.

\section{PSD estimation in beamspace}
\label{sec:proposed}
\begin{figure}[tb]
  \begin{center}
    \includegraphics[keepaspectratio=true,width=0.33\textwidth]{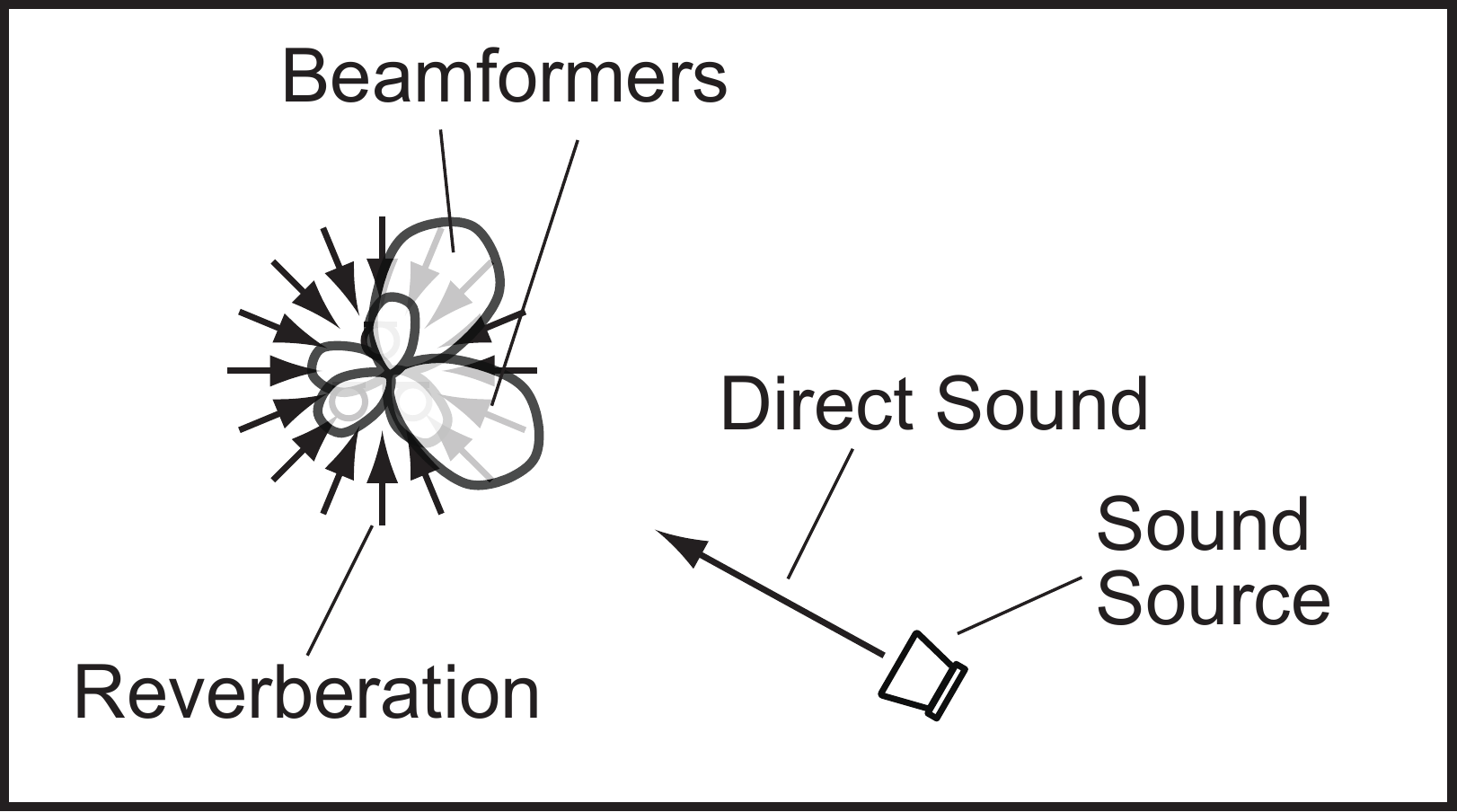}
  \end{center}
\vspace*{-1.0\baselineskip}
  \caption{PSD estimation in beamspace for estimating DRR. Two different beamformers are applied to the microphone array observation to create the beamspace for estimating the DRR.}
  \label{fig:PSDinBeamSpace.pdf}
\end{figure}
Assume two beamformers which have different directivity patterns are applied to the microphone array observation as shown in Fig. \ref{fig:PSDinBeamSpace.pdf}. According to (\ref{eq:BF_out_PSD_isotropic}) the output PSD of the two beamformers can be formulated in a matrix form given by (\ref{eq:simultaneous_eq}).
\begin{align}
&\underbrace{\left[\begin{array}{c}
P_{\mathrm{BF},1}(\omega)\\
P_{\mathrm{BF},2}(\omega)
\end{array}\right]}_{\mathbf{P}_{\mathrm{BF}}(\omega)}
=
\underbrace{
\left[\begin{array}{cc}
G_{1,\Omega_{\mathrm{D}}}(\omega) &  \int_{\Omega}G_{1,\Omega}(\omega)d\Omega\\
G_{2,\Omega_{\mathrm{D}}}(\omega) & \int_{\Omega}G_{2,\Omega}(\omega)d\Omega
\end{array}\right]}_{\mathbf{G}(\omega)}
\underbrace{\left[\begin{array}{c}
P_{\mathrm{D}}(\omega)\\
\overline{P}_{\mathrm{{R}}}(\omega)
\end{array}\right]}_{\mathbf{P}_{\mathrm{cmp}}(\omega)}
\label{eq:simultaneous_eq}
\end{align}
Because the elements in $\mathbf{P}_{\mathrm{BF}}(\omega)$ and $\mathbf{G}(\omega)$ are known \textit{a priori}, the PSD of the direct sound and the reverberation can be estimated by solving the simultaneous equation
\begin{align}
\hat{\mathbf{P}}_{\mathrm{cmp}}(\omega)=\mathbf{G}^{-1}(\omega)\mathbf{P}_{\mathrm{BF}}(\omega),
\end{align}
where $\hat{\cdot}$ denotes an estimated value. The DRR of each sound source is estimated by substituting $P_{\textrm{D}}(\omega)$ and $\overline{P}_{\textrm{R}}(\omega)$ in (\ref{eq:DRR}) by the estimated values found in $\hat{\mathbf{P}}_{\mathrm{cmp}}(\omega)$.


The previous method invented by the authors \cite{hioka_2012IWAENC} also uses two beamformers however their beampatterns have to be identical while their mainlobes are pointing in different directions. It is obvious that the previous method is solving (\ref{eq:simultaneous_eq}) given
\begin{align}
\int_{\Omega}G_{1,\Omega}(\omega)d\Omega&=\int_{\Omega}G_{2,\Omega}(\omega)d\Omega:=\int_{\Omega}G_{\Omega}(\omega)d\Omega\\
G_{1,\Omega_{\mathrm{D}}}(\omega)&\neq G_{2,\Omega_{\mathrm{D}}}(\omega)\label{eq:previous_condition2},
\end{align}
which is a specific case of PSD estimation in beamspace.

 Then the PSD of the direct sound is estimated from the difference of the output PSD of these beamformers
\begin{align}
\hat{P}_{\mathrm{D}}(\omega)=\frac{P_{1}(\omega)-P_{2}(\omega)}{G_{1,\Omega_{\mathrm{D}}}(\omega)-G_{2,\Omega_{\mathrm{D}}}(\omega)},\label{eq:PSD_direct_estimate}
\end{align}
whereas the PSD of the reverberation is calculated by the following subtraction of the estimated $P_{\mathrm{D}}(\omega)$ from the PSD of the microphone observation:
\begin{align}
\hat{P}_{\mathrm{R}}(\omega)=P_{X}^{(m)}(\omega)-\hat{P}_{\mathrm{D}}(\omega),
\end{align}
where $P_{X}^{(m)}(\omega)$ is derived by
\begin{align}
P_{X}^{(m)}(\omega)&=E[|X^{(m)}(\omega,t)|^{2}]_{t}.
\end{align}

\begin{figure}[tb]
\begin{center}
\subfigure[Ambient]{
\includegraphics[width=0.38\textwidth]{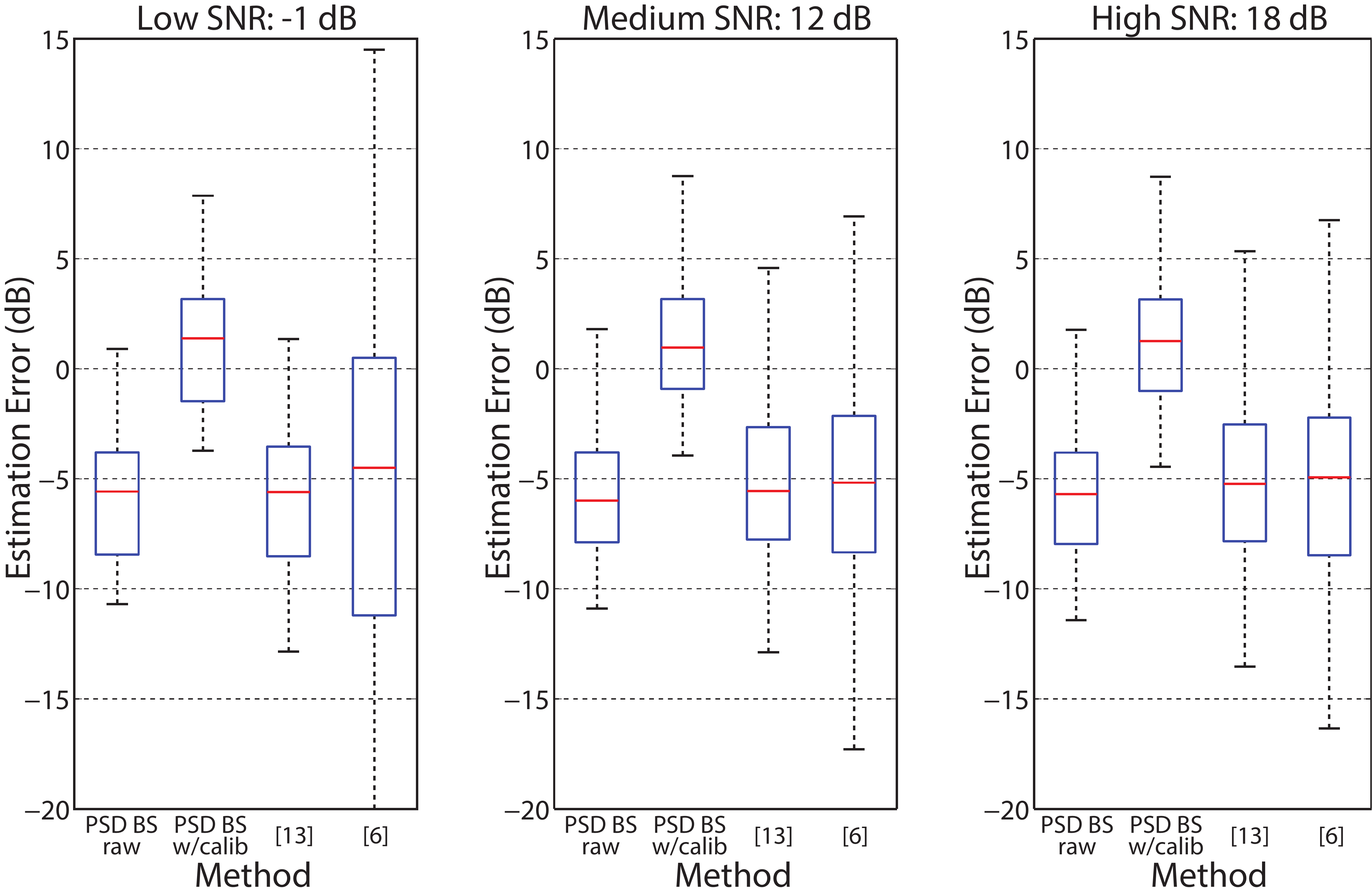}
}
\subfigure[Fan]{
\includegraphics[width=0.38\textwidth]{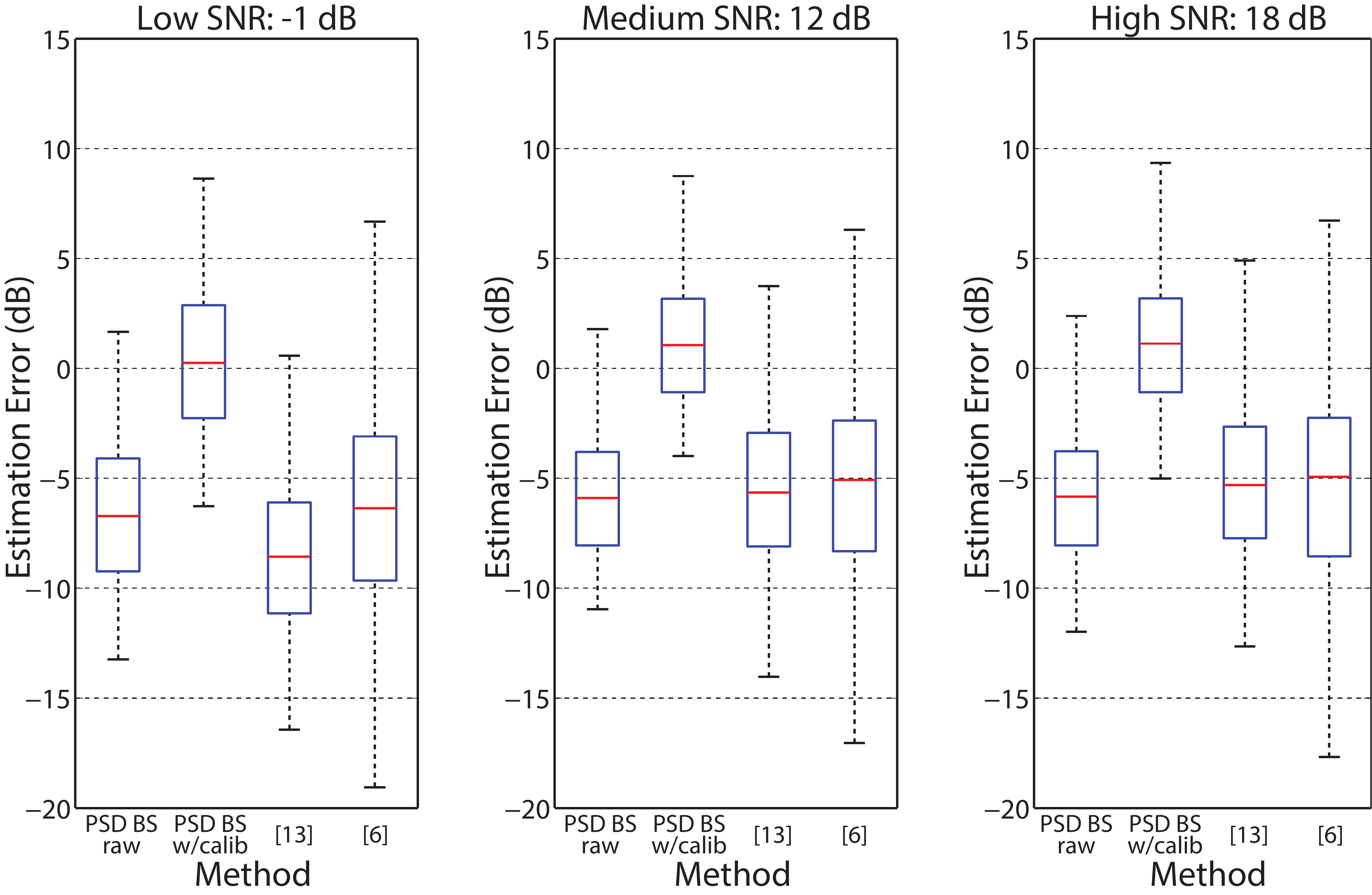}
}
\subfigure[Babble]{
\includegraphics[width=0.38\textwidth]{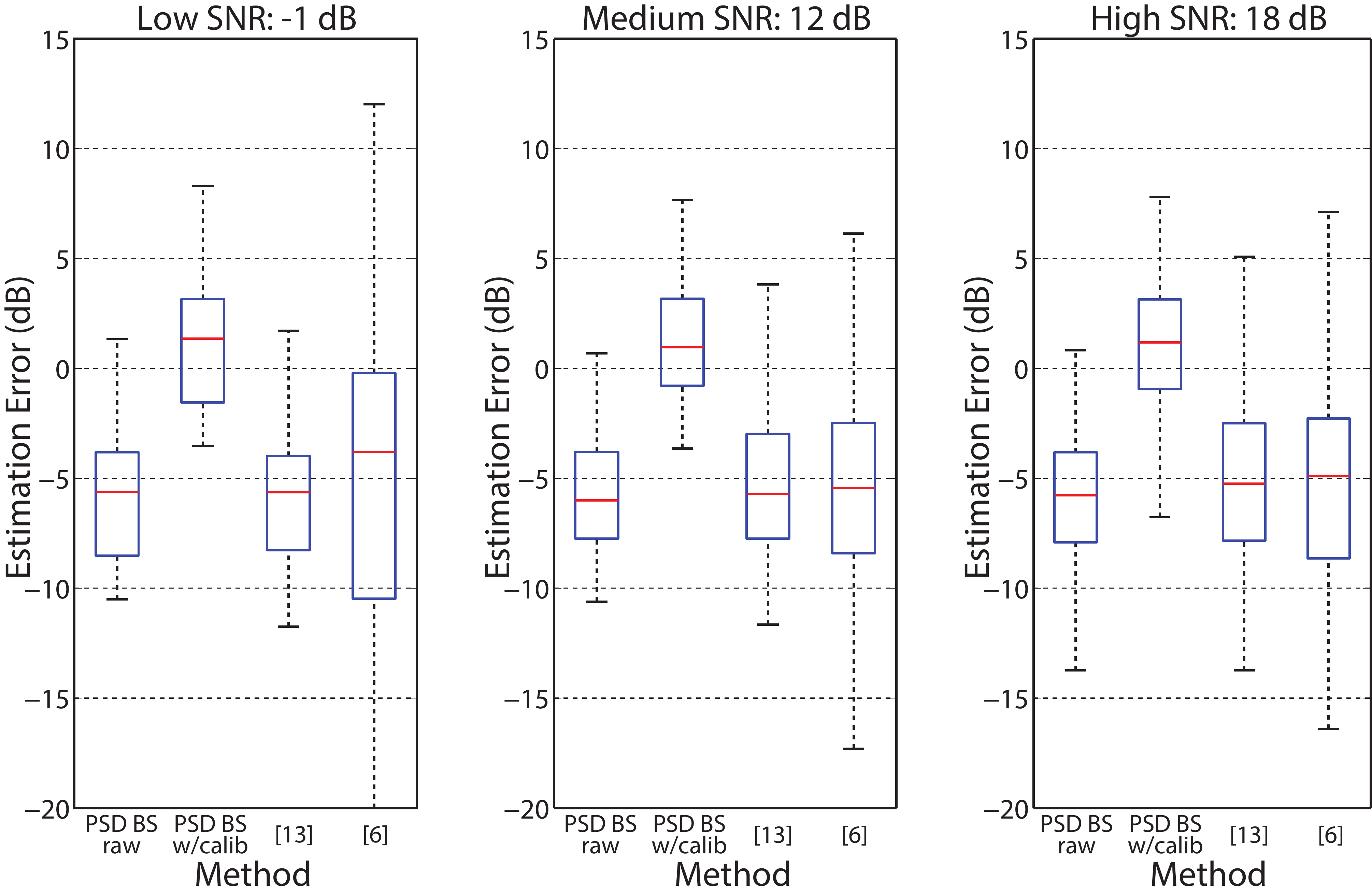}
}
\caption{Comparison of estimation error for different types of noise provided in the ACE speech corpus.}\label{fig:errorBYntype}
\end{center}
\end{figure}
\section{Performance evaluation}
The performance of the proposed method was evaluated along with that of the authors' previous methods \cite{HIOKA2011IEEE,hioka_2012IWAENC} using the ACE Challenge corpus \cite{Eaton2015a}. Among various microphone array configurations provided in the corpus, the 3-channel mobile microphone array (right-angled triangular configuration) was selected for the evaluation. The DOA of the direct sound $\Omega_{\mathrm{D}}=\{\theta_{\mathrm{D}},\phi_{\mathrm{D}}\}$ was estimated by the steered beamformer based method \cite{brandstein2001microphone} with a delay-and-sum beamforming \cite{Johnson_arraybook} used for the beamformer. Since speech signals are generally nonstationary, a voice activity detection (VAD) was applied for extracting frames which include reasonable amount of speech components, then only the extracted frames were used for calculating the expectation in the PSD estimation. The VAD was realised by simply selecting the frames of the short time Fourier transform of a microphone observation where the power of the signal in the frames exceeded a certain threshold, which was determined based on the level of stationary noise estimated by \cite{Niwa_IWAENC2014}.

Two delay-and-sum beamformers whose mainlobes being pointed in $\Omega_{1}=\{\theta_{\mathrm{D}},\phi_{\mathrm{D}}\}$ and $\Omega_{2}=\{\theta_{\mathrm{D}} + \frac{\pi}{3},\phi_{\mathrm{D}}\}$, respectively, were used for PSD estimation in beamspace. The signals (sampled at $16$ kHz \cite{Eaton2015a}) were analysed by the short-time Fourier transform whose frame size was $512$ samples. The frame was shifted by half the frame size (i.e. $256$ samples). The spatial resolution for the DOA estimation was $\frac{\pi}{72}$ rad for the azimuth and $\frac{\pi}{60}$ rad for the zenith angles.
Since, in the ACE Challenge, an opportunity was given to calibrate a method using a small dataset (i.e. Dev dataset \cite{Eaton2015a}), a calibration was introduced to the proposed method in order to compensate biases seen in the estimation. The biases would have been caused by the difference of the definitions of DRR between the ACE Challenge corpus and the proposed method. From the model introduced in Fig. \ref{fig:impulse_response.pdf}, all the early reflections were incorporated into the reverberation in the proposed method whereas some \textit{very} early reflections were accounted to be the direct sound in the corpus \cite{Eaton2015a}. The calibration simply biased all raw estimated DRR by a constant value, which was calculated by taking the average of the estimation errors given in the test with the Dev dataset. 

Fig. \ref{fig:errorBYntype} shows the comparison of the error distribution of the estimated DRR for different types and levels of noise in the tested acoustical environment provided in the full corpus set (i.e. Eval dataset \cite{Eaton2015a}). By looking at the extent of the distribution of the estimation errors (height of the box plots), the proposed method is more accurate compared to the previous methods. It is revealed that the method with coherence based PSD estimation \cite{HIOKA2011IEEE} suffers from very large variance of the estimation error compared to the methods based on PSD estimation in beamspace. Presumably the error in \cite{HIOKA2011IEEE} would be mainly caused by the fact that the method could not distinguish the reverberation arriving from the same direction as that of the direct sound \cite{hioka_2012IWAENC}. The results also show the bias calibration contributed to improve the estimation accuracy.

In order for further investigating the performance of the proposed method in different acoustical environment, a distribution of the estimation error in each acoustical environment included in the Eval dataset (2 different distances in 5 different rooms: Office 2, Meeting Rooms 1\&2, Lecture rooms 1\&2 \cite{Eaton2015a}) is shown in Fig. \ref{fig:errorBYenv}. From the result it is clearly seen that both the proposed and previous methods have been affected by the change of the acoustical environment. One of possible causes of this adverse effect would be the poor accuracy of the pre-estimated DOA since both the proposed and previous methods strongly rely on the DOA of the speaker. This fact should be a good guidance for future study to develop a more practically sound DRR estimation method which should be less dependent on the DOA information.

\begin{figure}[tb]
\begin{center}
\includegraphics[width=0.36\textwidth]{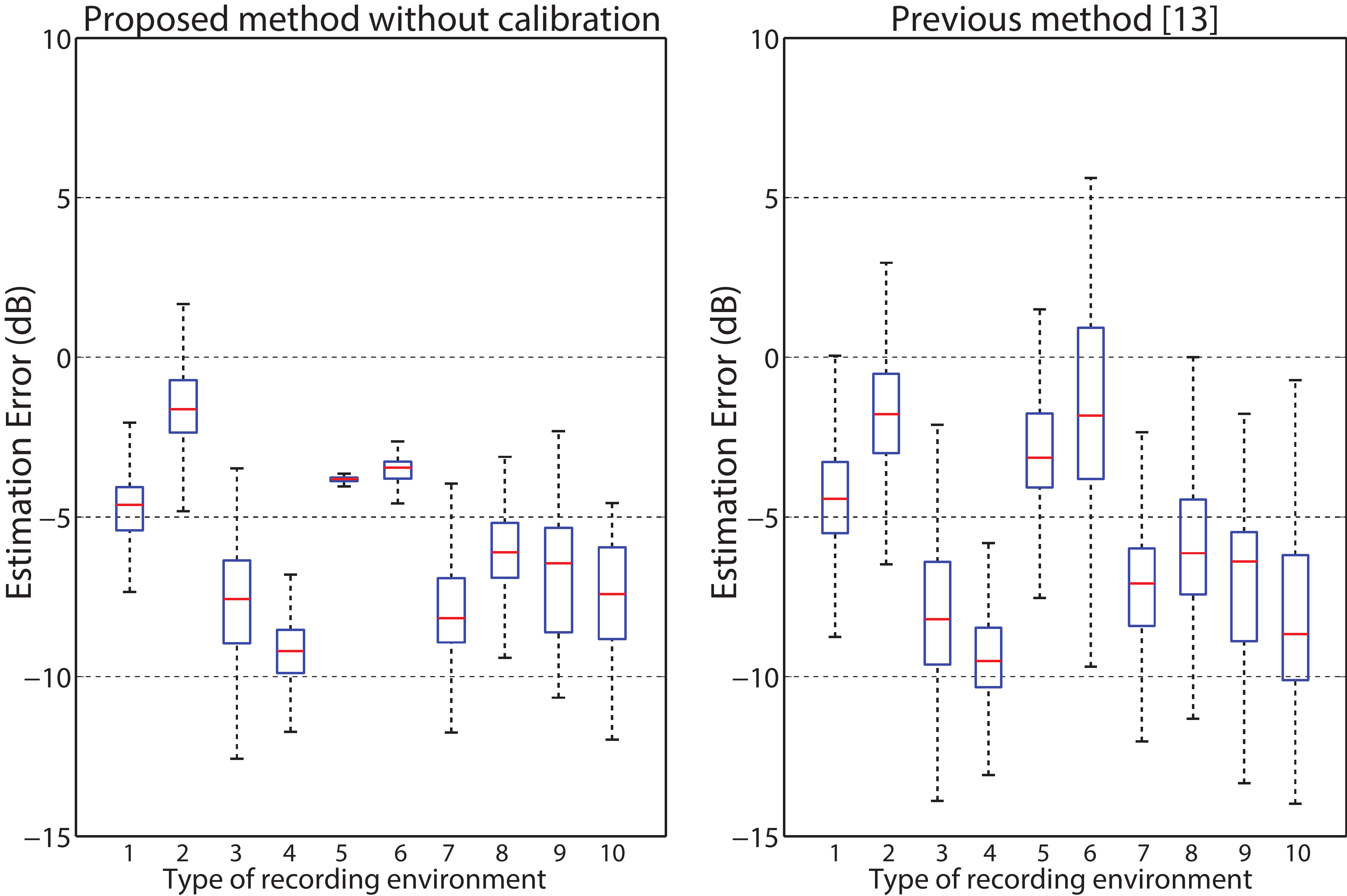}
\caption{Distribution of the estimation error for different experimental setups provided in the ACE corpus.}\label{fig:errorBYenv}
\end{center}
\end{figure}

\section{Conclusion}
A method for estimating DRR from speech signals has been proposed. The method estimates the PSD of the direct sound and reverberation using the output signal of two beamformers applied to a microphone array observation. The estimated PSD's are used for calculating the DRR. The proposed method was evaluated by experiments using the speech corpus of the ACE Challenge. Nevertheless the proposed method suffered from its estimates being biased, which may be compensated by some preliminary calibration, it was revealed that the proposed method provides more accurate estimates of the DRR compared to the authors' previous methods. Further improvement should be sought to make the proposed method more robust to the variation of the acoustical environment and less dependent on the estimated DOA of the speech signal.  
\clearpage
\bibliographystyle{IEEEtran}
\bibliography{refs}
%
%
%
%
%
%
%
%
%

\end{sloppy}
\end{document}